# On the 27-day Variations of Cosmic Ray Intensity in Recent Solar Minimum 23/24

R. Modzelewska[a] and M.V. Alania[a, b]

*(a) Institute of Math. and Physics, Siedlce University, 08-110 Siedlce, Poland*
*(b) Institute of Geophysics, Tbilisi State University, Tbilisi, Georgia*

*ABSTRACT*
We have studied the 27-day variations and their harmonics of the galactic cosmic ray (GCR) intensity, solar wind velocity, and interplanetary magnetic field (IMF) components in the recent prolonged solar minimum 23/24. The time evolution of the quasi-periodicity in these parameters connected with the Sun's rotation reveals that their synodic period is stable and is ≈26-27 days. This means that the changes in the solar wind speed and IMF are related to the Sun's near equatorial regions in considering the differential rotation of the Sun. However, the solar wind parameters observed near the Earth's orbit provide only the conditions in the limited local vicinity of the equatorial region in the heliosphere (within $\pm 7°$ in latitude). We also demonstrate that the observed period of the GCR intensity connected with the Sun's rotation increased up to ≈33-36 days in 2009. This means that the process driving the 27-day variations of the GCR intensity takes place not only in the limited local surroundings of the equatorial region but in the global 3-D space of the heliosphere, covering also higher latitude regions. A relatively long period (≈34 days) found for 2009 in the GCR intensity gives possible evidence of the onset of cycle 24 due to active regions at higher latitudes and rotating slowly because of the Sun's differential rotation. We also discuss the effect of differential rotation on the theoretical model of the 27-day variations of the GCR intensity.

1. Introduction
Currently when the new activity of the 24th solar cycle has begun, it is a good occasion to discuss features of the whole solar minimum 23/24 (the minimum between cycles 23 and 24), which is a broadly investigated period because of its very peculiar nature (see, *e.g.*, McComas *et al.*, 2008; Smith and Balogh, 2008; Gibson *et al.*, 2011; Lee *et al.*, 2011, Dikpati, 2011). Fortunately, nowadays measurements of solar activity and solar wind parameters are effectively performed by a large number of operating worldwide ground-based observatories and space probes. The values of solar activity indices in the relatively long-lasting minimum 23/24 are appreciably different in comparison with the previous solar minima. The Sun was extremely quiet, with almost no sunspots on its surface (*e.g.*, Smith, 2011). This unusually long and deep solar minimum has been also detected in a variety of other solar cycle indices, such as solar radio flux and total solar irradiance (*e.g.*, Fröhlich 2009; Domingo *et al.*, 2009). Simultaneously, the observations at Wilcox Solar Observatory (WSO) reveal that the solar magnetic field is reduced over the whole surface of the Sun (Lee *et al.*, 2011), and the polar fields are about half of those observed during the previous minimum period (Hoeksema, 2009). The mean value of the strength *B* of the interplanetary magnetic field (IMF) in the period of 2007-2009 was record-low (≈2.5 nT) in comparison with the previous minimum epochs (≈4 nT in 1985-1987 and ≈3.4 nT in 1995-1997). The reason for the decrease in IMF in this minimum 23/24 is still under discussion: it may be either due to less input from interplanetary coronal mass ejections (ICMEs) (Owens *et al.*, 2008) or a weaker input from solar polar magnetic flux (Cliver and Ling, 2011). Alternatively, Zhao and Fisk (2011) argued that actually the magnetic flux was the same in the two past minima, but the field strength was lower in this minimum 23/24 because it filled a larger area as a consequence of narrower streamer stalks. The consequences of these weaker solar magnetic fields were visible in changes of solar wind plasma flow through the interplanetary medium. Observations near the

Earth's orbit during the ending phase of solar cycle 23 demonstrated that the IMF strength and solar wind density were about 30% lower than in the minimum 22/23 and the momentum flux was about 38% lower (Lee *et al.*, 2011), whereas the solar wind speed remained unchanged in comparison to the previous solar minimum (Lee *et al.*, 2009b). At the same time the *Ulysses* (Balogh *et al.*, 1992; Bame *et al.*, 1992) out-of-the-ecliptic observations show comparable changes throughout the heliosphere, namely the reduction in the radial magnetic field by about 64%, dynamic pressure by about 22%, and thermal pressure by about 25% (Smith and Balogh, 2008; McComas *et al.*, 2008).

Generally active regions on the Sun's surface exhibit the latitudinal drift of the sunspot-generating zone in the course of the 11 year cycle (Carrington, 1858), which leads to the butterfly diagram (Maunder, 1904). It has been proposed that the extension of the meridional circulation and its speed determines the length of the solar cycle (Dikpati and Charbonneau, 1999). Dikpati (2011) reported that the meridional circulation pattern behaved distinctly differently during solar cycles 22 and 23. Particularly, the Sun's surface plasma flow was poleward all the way up to the pole and seemed to make up one large flow cell that persisted during the major part of solar cycle 23. During solar cycle 22, however, the poleward surface flow ended at around 60° latitude, beyond which it became equatorward making up a two-cell flow pattern. Dikpati (2011) further noted that the maximum average flow speed at the surface was not different during solar cycles 22 and 23. Altrock (2010) showed that emission of active regions in the peculiar solar cycle 24 began similarly to previous cycles, although at a 40% slower rate.

Comparison of observations from the first three orbits of *Ulysses* demonstrated that the 3-D structure of the solar wind varies dramatically over the solar cycle (McComas *et al.*, 2006). They suggest that this difference might be a regular feature of the full ≈22-year Hale cycle in considering the polarity changes of solar magnetic fields. Lario and Roelof (2007) reported that the long-lived and well defined ≈26-day recurrent intensity enhancement is observed in the minimum epoch of solar activity when $A>0$ (the north pole had positive polarity) during the first (1992 - 1994) southern excursion of *Ulysses*. In contrast to this, during the third (2005 - 2007) southern excursion of *Ulysses*, taking place also in low activity conditions but when $A<0$, observations showed more variable structure in the solar wind stream. Dunzlaff *et al.* (2008) suggest that this might be caused by the difference in the coronal hole structures between cycles 22 (1986-1996) and 23 (1996-2008); a large, stable coronal hole structure was present during cycle 22 but not in cycle 23. The polar coronal hole disappeared during the declining phase of cycle 23 (Kirk *et al.*, 2009) and a part of the coronal hole structure existed in the equatorial region (Abramenko *et al.*, 2010).

The most probable source of the differences between solar cycles 22 and 23 is the dynamics of the Sun's interior which, through the dynamo process, influences the generation and evolution of the Sun's global magnetic fields. Consequently, the modulation of galactic cosmic rays (GCR) in solar cycles 22 and 23 has recently been explained by the differences in evolution pattern of coronal holes connected with the dynamo-generated large-scale magnetic fields (Dikpati, 2011). On the other hand Zhao and Fisk (2011) reported that in the recent minimum 23/24 the fundamental physical process that accelerates the solar wind is unchanged and the global model for the behavior of the IMF remains valid.

Modulation of the GCR intensity in the minimum 23/24, because of its very unusual character, is also a broadly investigated subject (*e.g.*, Heber *et al.*, 2009; McDonald *et al.*, 2010; Schwadron *et al.*, 2010). Recently Paouris *et al.* (2012) studied the GCR modulation in the minimum 23/24 in relation to solar activity indices and heliospheric parameters, such as the sunspot number, interplanetary magnetic field, CME index, and the tilt angle of the heliospheric current sheet.

It is well known that the diffusion coefficient $K$ of GCR particles depends, among other parameters, on the magnitude $B$ of the IMF as $K \propto 1/B$. Therefore, the unusually weak IMF in the minimum 23/24 should lead to relatively high parallel diffusion coefficient, which naturally causes higher GCR intensities that were observed by neutron monitors (*e.g.*, Moraal and Stoker, 2010) and space probes (*e.g.*, Mewaldt *et al.*, 2010) as well.

During solar minima, transient disturbances in the interplanetary space are infrequent and the regular IMF is well established. Therefore, it is relatively easy to distinguish the quasi-periodic variations caused by the Sun's rotation. For the observer on the Earth the recurrence period of such variations (the synodic period) is about 27-28 days considering the differential rotation of the Sun near the equatorial regions.

The 27-day variations of the GCR intensity are generally caused by the (heliographic) longitudinal asymmetry of the electro-magnetic conditions in the heliosphere during one solar rotation. The variations are caused by the inhomogeneous distribution of active regions on the Sun's surface and depend on the lifetime of this asymmetry in the interplanetary space (Dorman, 1961). This topic, first announced by Forbush (1938), has been discussed extensively (for a review, see *e.g.*, Richardson, 2004).

Generally, higher harmonics of the 27-day GCR variations (*e.g.*, 14 and 9 days) are related to simultaneous existence of several active longitudes (Alania and Shatashvili, 1974). Particularly, the period of ≈ 14 days comes mainly from two groups of active regions roughly $180°$ apart in longitude. This second harmonics was studied in detail, *e.g.*, by Mursula and Zieger (1996). Recently, Sabah and Kudela (2011) presented the power spectra of GCR time series in the period range of $T < 27$ days from measurements of neutron monitors and muon telescopes and discussed the connections between the higher harmonics, particularly the third harmonic (≈9 days), and the power spectra of the geomagnetic activity and interplanetary parameters.

In the recent minimum 23/24, the peculiarities in the 27-day recurrent variations were clearly manifested in a variety of cosmic ray counts detected by neutron monitors (*e.g.*, Alania *et al.*, 2010) and space probes (*e.g.*, Leske *et al.*, 2011). The magnitude of the 27-day variations of the solar wind speed remains almost at the same level as in the previous minimum 22/23 (1995-1997), but the range of observed solar wind speed is larger with the significant second (≈14 days) and third (≈9 days) harmonics (Modzelewska and Alania, 2012). Recently, Modzelewska and Alania (2012) and Gil *et al.* (2012) suggested that the peculiarities in the amplitude of the 27-day variations of the GCR intensity and its dependence on the Sun's global magnetic field may be due to the large-scale structures of the solar wind speed and IMF with their stable longitudinal asymmetries.

In this paper we will further study the properties of the first (≈27 days), second (≈14 days) and third (≈9 days) harmonics of the 27-day variations of the GCR intensity, the solar wind velocity, and components of the IMF in the minimum 23/24. Also we will study their synodic quasi-periodicity connected to the Sun's rotation and the powers of these quasi-periodic variations. Based on spectral and wavelet analysis methods, we will investigate the temporal variations of the synodic periodicity in the selected solar wind parameters. By combining these with the analysis of GCR intensity variations, possible evidence for the onset of new 24th solar cycle will be proposed.

2. Experimental Data
Generally in the activity minimum, sunspots exist both in lower latitudes, belonging to the old cycle, and in middle latitudes, belonging to the new cycle. Therefore, it is interesting to know how different latitudinal locations of the longitudinal asymmetry in sunspot distributions and the solar wind properties affect the quasi-periodic variations of the GCR intensity owing to the existence of the Sun's differential rotation. For this purpose we study the synodic periodicity (period *T*) connected to the Sun's rotation and its power *P* in the GCR intensity and selected solar wind parameters, using spectral and wavelet analysis methods.

To reveal the synodic period *T* in the GCR intensity connected to the Sun's rotation, we analyze the data obtained by the Kiel neutron monitor [http://www.nmdb.eu]. The solar wind velocity V and $B_x$ and $B_y$ components of the IMF are taken from the OMNIWEB database [http://omniweb.gsfc.nasa.gov/ow.html]. These data cover the period of 2007-2009, corresponding to the 2367-2406 Bartel's rotation (BR) periods.

To study the temporal evolution of the 27-day variations of the GCR intensity and similar quasi-periodic variations in other parameters, we have divided the interval of 2007-2009 (2367-2406 BR periods) into three periods; I: BR 2367-2388, II: BR 2389-2395, and III: BR 2396-2406. This division is motivated by different character of the quasi-periodic variations of the GCR intensity; the 27-day variations of the GCR intensity are stable in period I, vanish in period II, and in period III they showed longer periods ($T \approx 34$ days). We then study the temporal evolution of the first ($\approx$27 days), second ($\approx$14 days), and third ($\approx$9 days) harmonics of the synodic periodicity (related to the Sun's rotation) in the GCR intensity and solar wind parameters during the minimum 23/24 (2007-2009). Long-term changes in the GCR intensity variations caused by the solar rotation in the period 1953-1994 were studied in *e.g.*, Basilevskaya *et al.* (1995).

To derive the period *T* in the analyzed time series, the method of power spectrum density was used. This method decomposes the time series in terms of components with different frequency ($\omega$) or period (*T*) (*e.g.*, Otnes and Enochson, 1972; Press *et al.*, 2002):

$$\psi(\omega) = \int_{-\infty}^{\infty} R(t) e^{-i\omega t} dt, \quad \omega = 2\pi f = \frac{2\pi}{T} \tag{1}$$

where *R(t)* is the autocorrelation function. In a discrete case we have

$$\psi(\omega) = \Delta t \sum_{r=-N}^{N} W \cdot R(r) e^{-i\omega r \Delta t} \tag{2}$$

where $R(r) = \frac{1}{N-r} \sum_{i=1}^{N-r} x_i x_{i+r}$ is the autocorrelation function and *W* is the window function (we use Parzen's window function). The power *P* of each period/frequency is calculated as $P = |\psi(\omega)|^2 \text{Hz}^{-1}$. To improve statistical reliability, the frequency components are calculated based on the time series consisting of 10 BR periods (270 days). The temporal changes of the recognized synodic periodicity are examined by removing and adding one BR period in the analyzed time series. To show the quasi-periodic character of the analyzed data we have fitted the time series by Fourier series truncated at order *m*, by the least square method (*e.g.*, Gubbins, 2004):

$$f(t) \approx a_0 + \sum_{j=1}^{m} (a_j \cos(j \cdot \omega t) + b_j \sin(j \cdot \omega t)), \quad m < \frac{N}{2} \tag{3}$$

where

$$a_0 = \frac{1}{N} \sum_{i=1}^{N} Y_i, \quad a_j = \frac{2}{N} \sum_{i=1}^{N} Y_i \cos(j \cdot X_i), \quad b_j = \frac{2}{N} \sum_{i=1}^{N} Y_i \sin(j \cdot X_i). \tag{4}$$

In the top panel of Figure 1(a) we show temporal variations of daily GCR intensity measured by the Kiel neutron monitor, and the approximations by the sum of the first (27 days), second (14 days), and third (9 days) harmonics according to Equations (3) and (4). Similar plots are made for the solar wind velocity **V** (Figure 1 (b)) and $B_x$ (Figure 1 (c)) and $B_y$ (Figure 1 (d)) components of the IMF. We do not consider contributions from the $B_z$ component in the magnitude $B = \sqrt{B_x^2 + B_y^2 + B_z^2}$ of the IMF due to its small values. Indeed, for the considered period of 2007-2009 (2367-2406 BR), the average values are $B_z^2$=0.52 nT$^2$, $B_x^2$=3.39 nT$^2$, and $B_y^2$=3.64 nT$^2$. Consequently, the contributions from the $B_z$ component is less than 10% of those from the $B_x$ and $B_y$ components. Figures 1(a)-1(d) clearly show the recurrent variations of ≈27-day period in cosmic ray counts detected by neutron monitors and in the solar wind parameters.

The middle panels of Figures 1(b)-1(d) show that the synodic period $T$ seen in the solar wind speed and components of the IMF is ≈26-27 days and is stable in all the analyzed three periods. The solar wind parameters we use were obtained by the spacecraft located at the Lagrangean point L1 and only reflect the wind conditions limited in the equatorial region of the heliosphere extending only up to ±7° latitudes. Therefore, it is not possible to derive direct relationship between the solar wind parameters at higher latitudes and the observed values around the equatorial region. On the other hand, the GCR intensity variations measured by neutron monitors reflect the conditions both in the limited local surroundings in the equatorial region, but in the global 3-D heliosphere covering also higher latitude regions.

Because of very peculiar behavior of the observed periodicity in the GCR intensity in periods II and III, we have carried out more precise analyses with the time resolution of one day instead of 1 BR. Namely, we have handled daily data series as follows: (1) First of all, the method of power spectral analyses was applied to data series $\{x_1, x_2, \ldots, x_n\}$; to obtain reliable results the length of the data series was taken to be 270 days (consisting of 10 BR periods) as mentioned above. The period $T$ of the component and its power $P$ with the 95% confidence level are computed. (2) Then we calculate the same parameters for the next series of data $\{x_2, x_3, \ldots, x_{n+1}\}$, i.e., we omit the data of the beginning day and add one extra day in the time series, and period $T$ and power $P$ are calculated. This process is repeated and we obtain $T$ and $P$ with the time resolution of one day.

We found that the synodic period $T$ of the GCR intensity (middle panel in Figure 1(a)) remains almost constant, ≈26-27 days, being in good agreement with similar changes in the solar wind velocity **V** and $B_x$ and $B_y$ components of the IMF for period I. In period II the stability of the synodic periodicity is weakened; one can see reduction in the power of the synodic periodicity. Finally in the last part of period II and in period III, we observe gradual increase in the period $T$ from 26-27 days up to 33-36 days. These findings demonstrate that: (1) the 27-day variations of the GCR intensity take place in the 3-D space of the heliosphere, and (2) the sources of such variations must include the contributions from active regions located at higher latitudes, rotating slower than the equatorial region by the differential rotation of the Sun.

The higher harmonics, namely the second (≈14 days) and third (≈9 days) harmonics of the synodic periodicity, are almost constant for all the parameters during the entire investigated period. This means that the sources of these harmonics have stable (heliographic) longitudinal and latitudinal distributions. The bottom panels of Figures 1(a)-1(d) present the temporal evolution of powers $P$ for the first ($P_{27}$), second ($P_{14}$) and third ($P_9$) harmonics of the 27-day variations of the GCR intensity, the solar wind velocity, and components of the IMF. For illustration we consider period I (BR 2367-2388). One can see very large power of the quasi-periodic variations connected to the Sun's rotation with different delay times. Namely, the maximum in power of the 27-day variations of the solar wind velocity ($P_{27}(V)$) is observed ≈ 3-4 BRs before that of the GCR intensity ($P_{27}(GCR)$). The maximum in power of the IMF components ($P_{27}(B_x)$ and $P_{27}(B_y)$) precedes that of the GCR intensity by ≈1BR.

To consider whether the quasi-periodic changes in the solar wind velocity **V** and in the IMF components $B_x$ and $B_y$ are the causes of the similar GCR intensity variations (with different delay times), we computed the time delay which maximizes the correlation coefficient *r* between the power of the 27-day variations of the GCR intensity ($P_{27}$ (GCR)) and the same power of all the analyzed parameters ($P_{27}$ (V), $P_{27}$ ($B_x$), and $P_{27}$ ($B_y$)). To study precisely the delay time among these parameters, we fixed the temporal changes of the GCR intensity and shifted all the other parameters with respect to GCR by delay times in units of BR. Table 1 shows the results; the highest correlation coefficients *r* are between GCR and **V** (*r*= 0.92±0.05) with the delay time of 3 - 4 BR, between GCR and $B_x$ component of the IMF (*r*= 0.94±0.04) with the delay time of ≈ 0 BR, and between GCR and $B_y$ component of the IMF (r= 0.75±0.08) with the delay time of ≈1 BR, respectively.

To study the temporal changes of the periodicity in GCR connected with the Sun's rotation, we adopted the wavelet time-frequency spectrum technique. The wavelet technique is described in detail by Torrence and Compo (1998). We used the wavelet software available at the website [http://paos.colorado.edu/research/wavelets/software.html]. In our calculation we used the Morlet wavelet mother function. In Figures 2(a)-2(c) are presented the wavelet analysis of the GCR intensity measured by the Kiel neutron monitor for periods I, II, and III, respectively. The presented results confirm that the 27-day variations of the GCR intensity were stable in period I (Figure 2 (a)), disappeared in period II (Figure 2 (b)), and in period III there was feeble indication of longer periods ≈ 32-35 days (Figure 2 (c)).

3. Theoretical Modeling

The main goal of this section is to compare theoretical predictions obtained from the numerical solutions to the transport equation with the experimental data of GCR intensity measured by neutron monitors. We model the period of 2007-2008 corresponding to BR numbers 2367-2388 (period I considered in the previous section). This choice may be justified due to the stable 27-day variations of the GCR intensity and similar quasi-periodic changes of the solar wind parameters and IMF.

To investigate theoretically the 27-day variations of cosmic rays we use a steady state, 3-D transport equation formulated by Parker (1965):

$$\nabla_i \cdot \left(K_{ij}^S \cdot \nabla_j f\right) - \left(v_{d,i} + V_i\right) \cdot \nabla_j f + \frac{1}{3}(\nabla_i \cdot V_i)\frac{\partial f}{\partial \ln R} = 0 \qquad (5)$$

where $f$ and $R$ are the distribution function and rigidity of cosmic ray particles, respectively; $V_i$ is the solar wind speed, $v_{d,i}$ is the drift velocity, and $K_{ij}^S$ is the symmetric part of the anisotropic diffusion tensor $K_{ij}$. The drift velocity is modeled as $\langle v_{d,i} \rangle = \partial K_{ij}^A / \partial x_j$ (Jokipii *et al.*, 1977), where $K_{ij}^A$ is the anti-symmetric part of the anisotropic diffusion tensor ($K_{ij} = K_{ij}^S + K_{ij}^A$) of GCR particles.

In the present paper we extend the model of our previous papers (Alania *et al.*, 2010, 2011) and consider the Sun's differential rotation as a simple kinematic indicator of dynamo-generated solar magnetic field and IMF. In this approach, the Sun's differential rotation is expressed by the angular velocity **Ω** that depends on the latitude as (Gibson, 1977): $\Omega(\theta) = 1 - 0.3\cos^2\theta$.

Based on the analysis of observed data of the GCR intensity, we ascribe its 27-day periodicity to the similar quasi-periodic changes (due to longitudinal asymmetry) of the solar wind velocity and IMF. In our model are included *in situ* measurements of the solar wind velocity, and the corresponding components of the IMF are obtained by solving a system of Maxwell's equations. The measured *in situ* radial speed was approximated by the first three terms of the Fourier series according to Equations (3) and (4) (Figure 3):

$V_r = V_0(1 + \alpha_1 \sin(1.65 + \varphi) + \alpha_2 \sin(2.6 + 2\varphi) + \alpha_3 \sin(0.4 + 3\varphi))$, $\alpha_1 = 0.09$, $\alpha_2 = -0.12$, $\alpha_3 = 0.12$ (6)

where $\varphi$ is the heliographic longitude. Using Equation (6) for *in situ* measurements of the solar wind velocity, we solve the system of Maxwell's equations:

$$\begin{cases} \dfrac{\partial \mathbf{B}}{\partial t} = \nabla \times (\mathbf{V} \times \mathbf{B}) & (7a) \\ \nabla \cdot \mathbf{B} = 0 & (7b) \end{cases}$$

where **V** is the solar wind velocity and **B** is the IMF. A complete description of our methodology of solving Maxwell's equations with variable solar wind speed and implementation of the solutions into Parker's transport equation was presented in our previous papers (Alania *et al.*, 2010, 2011).

Observations of solar wind parameters in the last minimum epoch of significantly low solar activity imply that the estimated parallel diffusion coefficient for the cosmic ray transport was considerably greater (*e.g.*, Moraal and Stoker, 2010; Mewaldt *et al.*, 2010). Therefore, to obtain better agreement between the theoretical model of the 27-day variations of the GCR intensity with the observed data for the last minimum epoch of solar activity (2007-2008), we increased the parallel diffusion coefficient by 40% ($\kappa_\parallel \approx 1.4 \times 10^{23}$ cm$^2$s$^{-1}$).

Recently it was demonstrated (Alania *et al.*, 2011) that the modulation parameter $\zeta$, which is proportional to the product of the solar wind velocity *V* and the strength *B* of the regular IMF ($\zeta \sim VB$), showed a remarkable negative correlation with the expected 27-day wave of the GCR intensity near solar minimum conditions. This is a hidden effect of the electric field in the cosmic ray transport, in general, for the convection-diffusion propagation of GCRs (Parker, 1965; Gleeson and Axford, 1967), and is also responsible for recurrent changes of the GCR intensity as well. Therefore, it is natural to examine this relationship between the product *VB* and the GCR intensity based on the observational data and the theoretical model in the whole recent solar minimum 23/24.

In Figures 4 (a) and 4(b) are presented the changes of both the observed (Kiel neutron monitor, Figure 4 (a)) and expected (Figure 4 (b)) GCR intensity (dashed line) and the product *VB* (solid line) during an average solar rotation for the minimum 23/24 in 2007-2008. The results of theoretical modeling for the 27-day variations of the GCR intensity with differential rotation of the Sun are in good agreement with the data of the Kiel neutron monitor. A remarkable negative correlation is also found between the product of *VB* and the 27-day wave of the GCR intensity, both in the theoretical model and in the observed data in 2007-2008. We confirm that an important role of the modulation effect, determined specifically by the product of the solar wind velocity *V* and the magnitude *B* of the IMF, is clearly manifested in the 27-day wave of the GCR intensity in the whole recent solar minimum 23/24.

Additionally, we investigated the correlation between the 27-day wave of GCR intensity and the solar wind velocity *V* or the magnitude *B* of the IMF, separately. One can see (Figures 5(a) and 5(b)) very high negative correlation between the solar wind speed and GCR intensity, both observationally (-78 %) and theoretically (-75%). Between the IMF *B* and GCR intensity (Figures 6(a) and 6(b)) this correlation is weaker (11% observationally, -56% theoretically). According to the theory of GCR modulation (Parker, 1965; Gleeson and Axford, 1967) the relationship between the product *VB* and the 27-day wave of the GCR intensity variations has an evident physical origin; this is a hidden effect of the electric field in the cosmic ray transport, as described above. Since *V* and *B* act together in this process, separation of this effect into solely the solar wind speed *V* or IMF *B* is impossible.

4. Conclusions
1.The quasi-periodic changes of the solar wind speed and IMF components related to the Sun's rotation demonstrate the existence of stable ≈26-27 day periodicity, which is in good agreement with the similar changes of the GCR intensity for period I (2007-2008, BR 2367-2388). However, there is some exception in the period of 2009, when the GCR intensity showed a gradual increase in the period from 26-27 days up to 33-36 days. We assign it to the nature of the formation of the 27-day variations of the GCR intensity, namely it takes place not only in the limited local surroundings of the equatorial region, but in the global 3-D space of the heliosphere including higher latitude regions. The observations of solar wind parameters reflect only changes in the limited local surroundings in the equatorial region.
2. Results of theoretical modeling for the 27-day variations of GCRs with differential rotation of the Sun are in good agreement with the observed data of the Kiel neutron monitor. We confirm that an important role of the modulation effect, determined specifically by the product ($VB$) of the solar wind velocity $V$ and magnitude $B$ of IMF, is clearly manifested in the 27-day wave of the GCR intensity in recent solar minimum 23/24.

**Figure captions:**

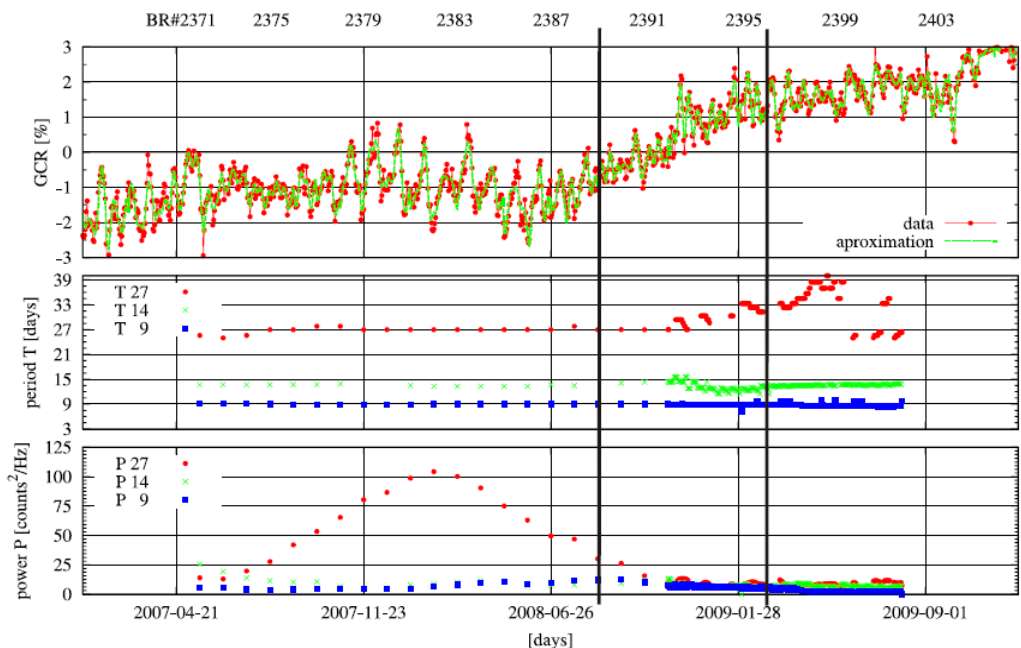

Figure 1 (a) Top panel: Temporal evolution of the daily GCR intensity in the recent solar minimum 23/24 measured by the Kiel neutron monitor (red dots connected by solid lines), and the approximation represented by the sum of the first (27 days), second (14 days), and third (9 days) harmonics (green dotted lines). Middle panel: Periods of GCR variations connected with the Sun's rotation; the first harmonic ($T \approx$ 27 days, red circles), the second harmonic ($T \approx$ 14 days, green crosses), and the third harmonic ($T \approx$ 9 days, blue squares). Bottom panel: Power *P* of the recognized periodicity; the first harmonic (red circles), the second harmonic (green crosses), and the third harmonic (blue squares). The vertical lines designate the boundaries of period I (BR 2367-2388), period II (BR 2389-2395), and period III (BR 2396-2406).

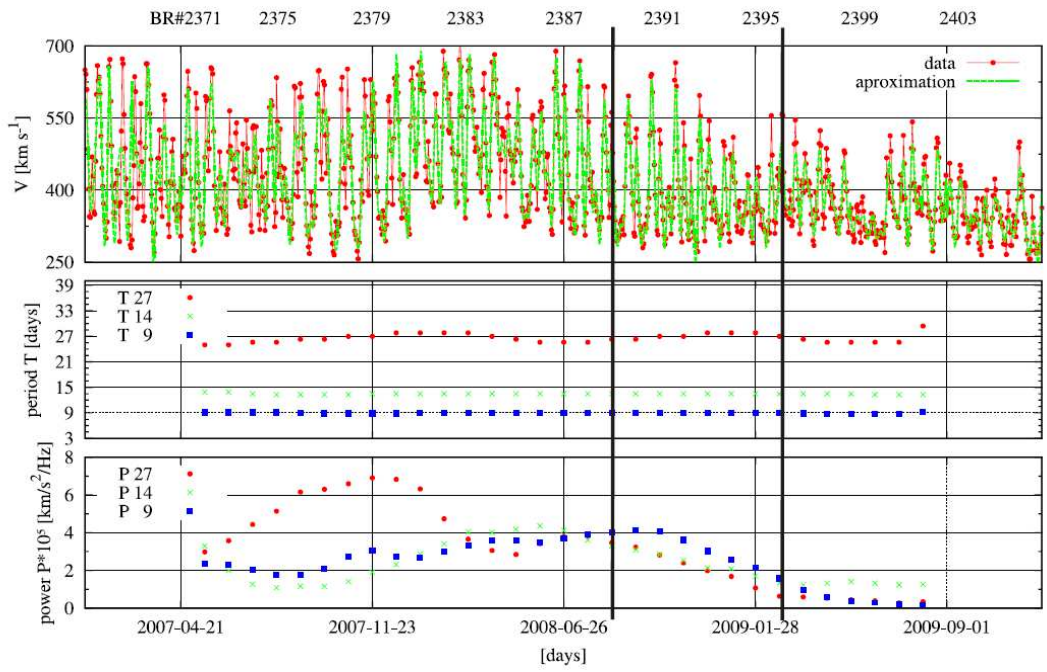

Figure 1 (b) The same as Figure 1 (a) but for the solar wind velocity *V*.

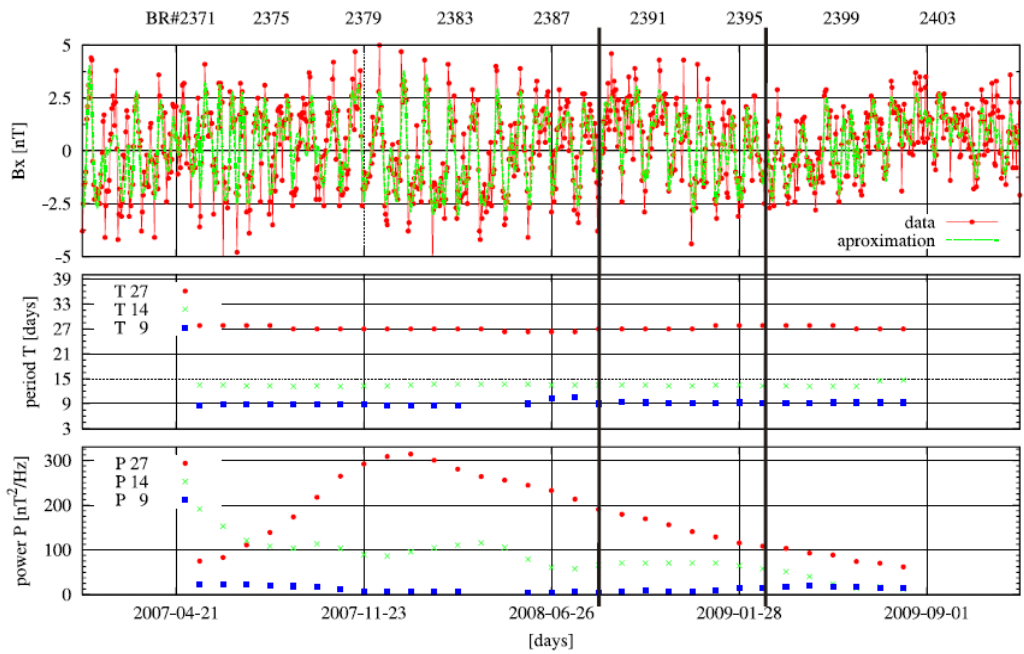

Figure 1(c) The same as Figure 1 (a) but for the $B_x$ component of the IMF.

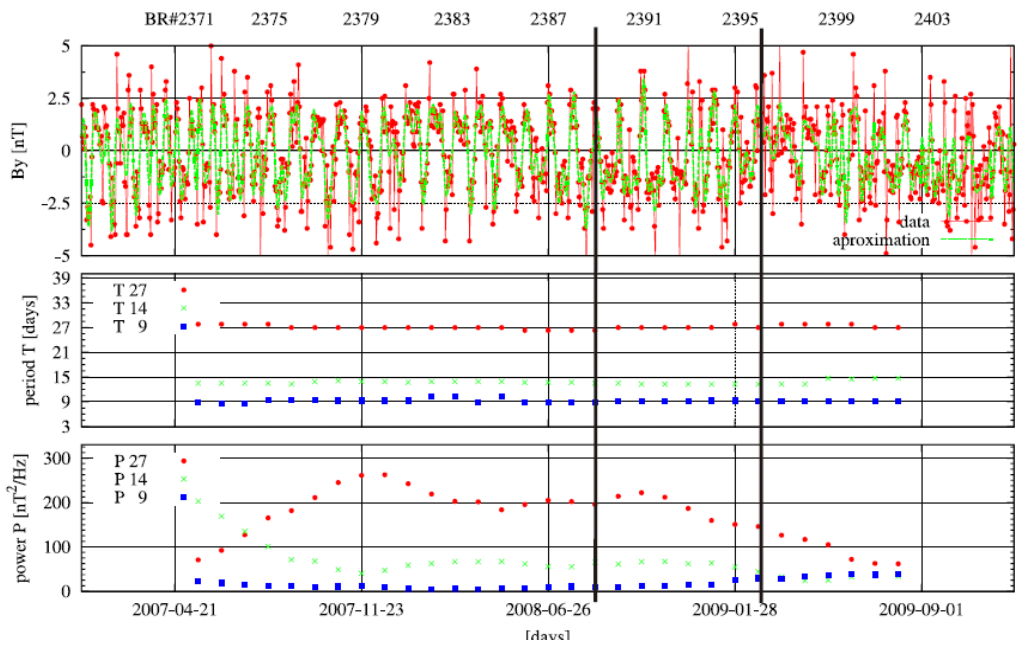

Figure 1 (d) The same as Figure 1 (a) but for the $B_y$ component of the IMF.

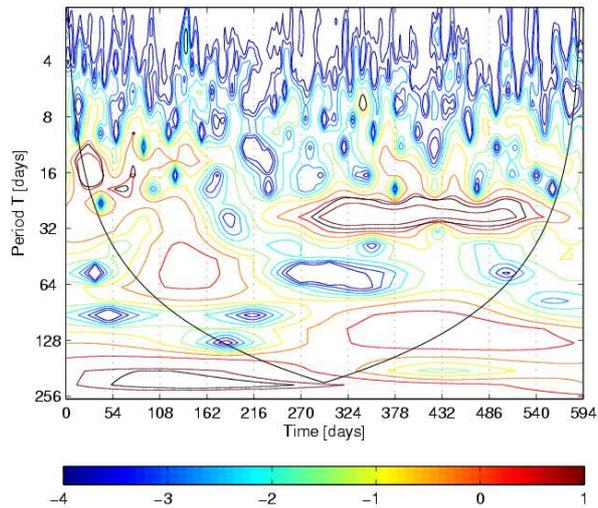

Figure 2 (a) Wavelet analysis of the GCR intensity measured by the Kiel neutron monitor for period I (BR 2367-2388).

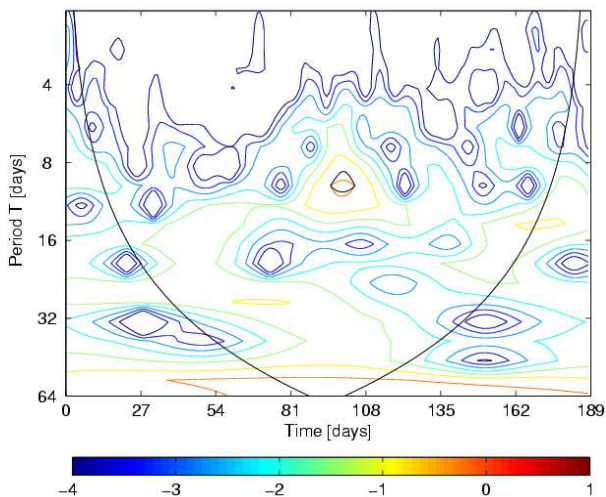

Figure 2 (b) Wavelet analysis of the GCR intensity measured by the Kiel neutron monitor for period II (BR 2389-2395).

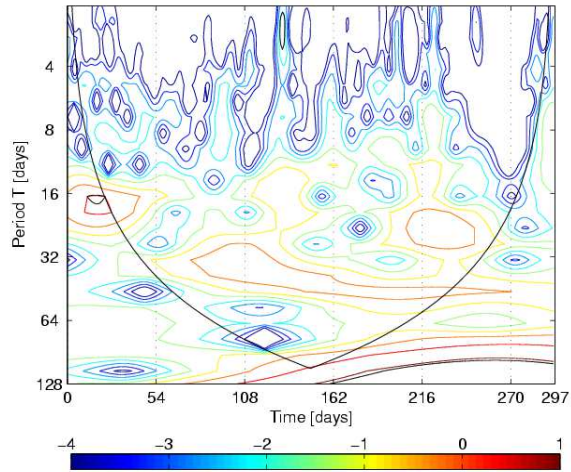

Figure 2 (c) Wavelet analysis of the GCR intensity measured by the Kiel neutron monitor for period III (BR 2396-2406).

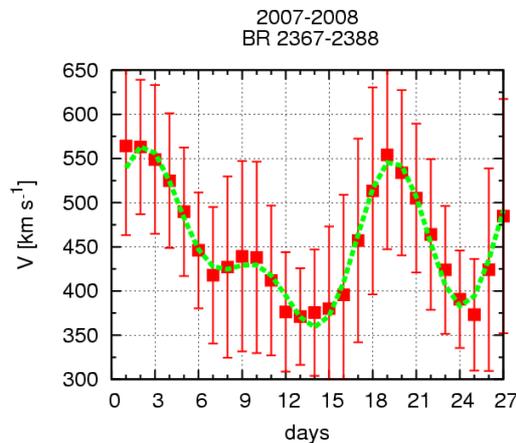

Figure. 3 Temporal changes of the solar wind speed (points with error bars) in 2007-2008 corresponding to BR 2367-2388 (period I) The fitting by the Fourier series with three terms is shown superimposed (green dotted lines).

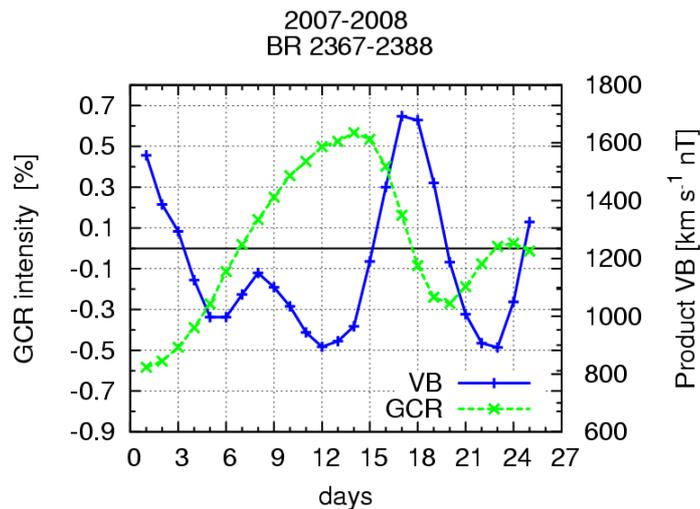

Figure 4 (a) Changes during one solar rotation in 2007-2008 corresponding to BR 2367-2388 (period I) of the observed GCR intensity at the Earth orbit (green dashed curve with crosses) and the product *VB* (blue solid curve with pluses).

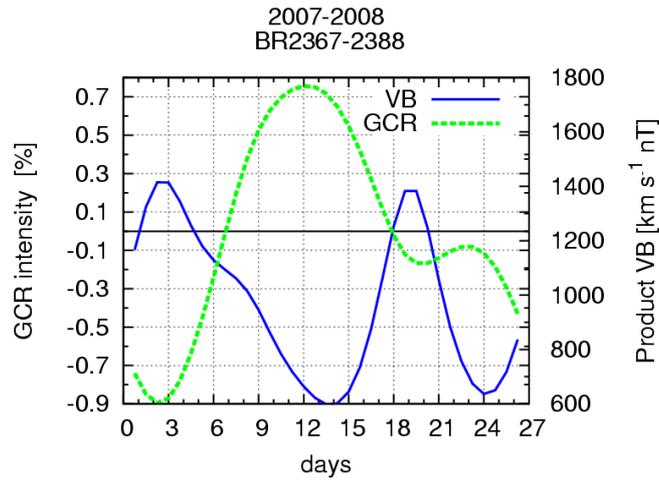

Figure 4 (b) Same as Figure 4(a) but for the GCR intensity of effective rigidity ≈10 GV at the Earth orbit (green dashed curve) and the product VB (blue solid curve) expected from the model.

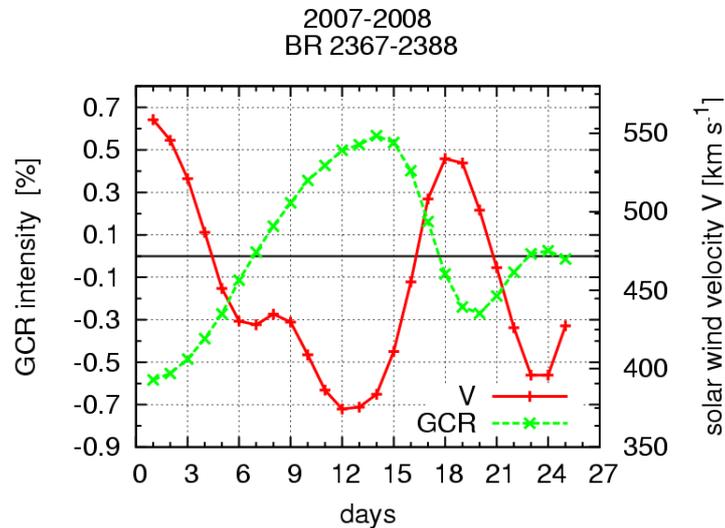

Figure 5 (a) Same as Figure 4(a) but for the observed GCR intensity (green crosses) and the solar wind speed *V* (red pluses).

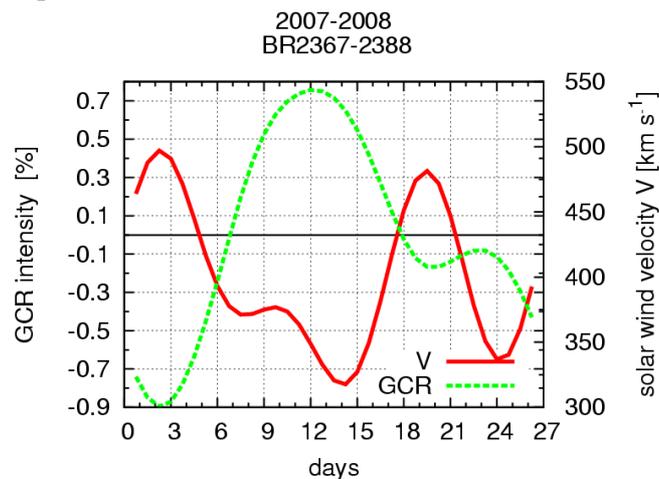

Figure 5 (b) Same as Figure 4(b) but for the GCR intensity of effective rigidity ≈10 GV at the Earth orbit (green dashed curve) and the solar wind speed *V* (red solid curve) expected from the model.

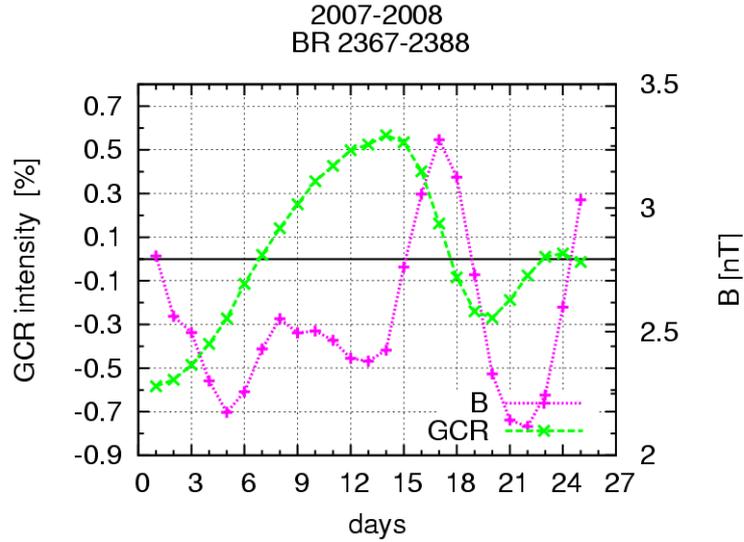

Figure 6 (a) Same as Figure 4(a) but for the observed GCR intensity at the Earth orbit (green crosses) and the magnitude $B$ of the IMF (purple pluses).

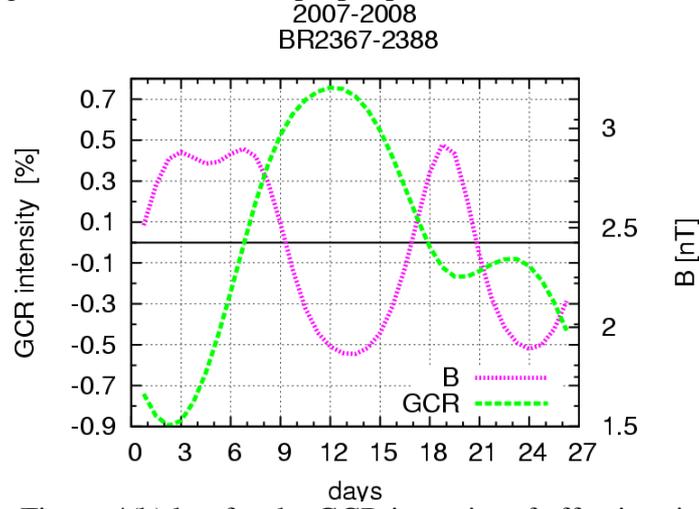

Figure 6 (b) Same as Figure 4(b) but for the GCR intensity of effective rigidity ≈10 GV at the Earth orbit (green dashed curve) and the magnitude $B$ of the IMF (purple dotted curve) expected from the model.

Table 1
The correlation coefficients, as a function of the delay time in BR, between the powers of the 27-day variations of the GCR intensity ($P_{27}$ (GCR)) and those of the solar wind velocity ($P_{27}$ (V)) and the $B_x$ and $B_y$ components of the IMF ($P_{27}$ ($B_x$), $P_{27}$ ($B_y$)).

| Delay time [BR] | 0 | 1 | 2 | 3 | 4 | 5 |
|---|---|---|---|---|---|---|
| $P_{27}$ (V)     | 0.42±0.11 | 0.66±0.09 | 0.87±0.06 | 0.92±0.05 | 0.92±0.05 | 0.87±0.06 |
| $P_{27}$ ($B_x$) | 0.94±0.04 | 0.92±0.05 | 0.84±0.07 | 0.72±0.09 | 0.55±0.10 | 0.36±0.11 |
| $P_{27}$ ($B_y$) | 0.72±0.09 | 0.75±0.08 | 0.74±0.08 | 0.66±0.09 | 0.51±0.11 | 0.33±0.12 |